# Immersive Virtual Reality in Learning Physics: A systematic review of opportunities and challenges

*Andrej Vidak[1], Iva Movre Šapić[1], Vjeran Gomzi[2], Vanes Mešić[3]

[1]Faculty of Chemical Engineering and Technology, University of Zagreb, Marulićev trg 19, 10000 Zagreb, Croatia
[2]Faculty of Electrical Engineering and Computing, University of Zagreb, Unska 3, 10 000 Zagreb, Croatia
[3]Faculty of Science, University of Sarajevo, Zmaja od Bosne 35, 71000 Sarajevo, Bosnia and Herzegovina

*Corresponding author email: avidak@fkit.unizg.hr

Immersive virtual reality (IVR) can heighten presence and enable active, embodied interaction in realistic 3D environments, which has been associated with potential benefits in physics education. This article presents a comprehensive review of immersive virtual reality implementation in physics education at both school and university levels. The analysis included 34 studies indexed in Scopus and ERIC, published between January 1, 2018, and June 1, 2025. Methodologically, this review followed a PRISMA approach. IVR implementations for physics learning were assessed, topic-aligned activity types were summarized, and associated opportunities and challenges were documented. Findings indicate that IVR can support physics learning by providing complementary visualizations, optimizing cognitive load, enabling haptic learning, saving time, and fostering collaborative inquiry. Conversely, the most frequent potential IVR challenges include discomfort, unreliable interaction, orchestration issues, relatively high costs, and significant extraneous cognitive load.

## 1. Introduction

Increasing technological progress has unlocked new methods for mastering phenomena that are traditionally difficult to conceptualize. For instance, modern technologies possess the profound potential to enrich our direct experience of scientific phenomena through immersive environments. A recent and influential classification of these computer-mediated realities is the xReality (XR) framework developed by Rauschnabel et al. (2022)[1]. According to this framework, when the real environment remains at least visually present within an experience, the result is augmented reality (AR); if the visual aspects of the real world are not present and are fully replaced by computer-generated content, the result is virtual reality (VR). Put simply, AR “represents a combination of real and virtual content that is displayed in real time,” whereas VR replaces the user’s visual access to the physical environment with synthesized imagery [2]. Within the xReality framework, assisted reality (AR with text or heads-up overlays) and mixed reality (AR with seamless, often occlusion-aware integration of complex virtual and real objects) are treated as subtypes of AR that differ primarily in the depth of integration. The authors further argue that “extended reality” should be avoided as an umbrella term because VR does not extend reality but visually replaces it; instead, XR is best interpreted as “xReality,” where the X serves as a placeholder for a family of reality formats that includes AR and VR, conceptualized along distinct continua.

Virtual reality (VR) can be further divided into two types: desktop virtual reality (DVR), and immersive virtual reality (IVR). DVR displays 3D scenes on standard 2D computer monitors and relies on a keyboard, mouse, or joystick for interaction. On the other hand, IVR employs a head-mounted display that occludes the physical world and presents a stereoscopic, computer-generated 3D environment to fully immerse users. IVR systems use sensors to detect the user's movements and adjust the view on the screen in real time. This allows users to move through a series of virtual spaces, linking their virtual movements and perspectives to the actual movements of their head and body. These technical features of IVR support a strong sense of presence – learners feel they are “in” the virtual environment and that events are really occurring [3]. In addition to entertainment, virtual reality is often used for training in professions such as astronauts and surgeons. It has made an important contribution to the development of science and education [4-6].

Modern high-end VR head-mounted displays (HMDs) such as the HTC Vive and Meta Quest deliver high immersion, unlike mobile VR options such as Google Cardboard or Samsung Gear VR. These premium HMDs feature advanced tracking and input accessories that precisely regulate immersion. They support six degrees of freedom (6 DOF) – three translations (left/right, up/down, forward/backward) and three rotations (roll, pitch, yaw) – enabling natural movement and interaction in virtual environments.

Leveraging these hardware capabilities, well-designed experiences at different immersion levels can make IVR a distinctive learning medium. In recent years, immersive virtual reality has created new opportunities for the education system and introduced new techniques and methods that have greatly facilitated and improved learning [7]. The ability to visualize theory and apply it in practice has facilitated understanding and brought people closer to concepts they could not previously imagine. This educational approach conveys important facts in a fun and simple way, making them easier for students to understand and presenting realistically potentially hazardous phenomena or costly environments.

Physics often deals with abstract, invisible processes that are difficult for many students to grasp. In physics education, virtual reality can help overcome this barrier by allowing students to visualize, simulate, and better understand these complex phenomena. With virtual reality, it is possible to completely overcome material limitations and scales, to see details that are not visible to the naked eye and to understand underlying physical processes [8]. For example, it is hard to imagine the existence of an electromagnetic field and to grasp what is happening within it. It is even more difficult to visualize fields based solely on the assumption that they are real. With virtual reality, electromagnetic fields can be visualized, making it easier to understand all the processes occurring inside them [9].

Another notable pattern in IVR-supported learning is the predominant use of Unity as the development platform, enabling advanced graphics and richly modeled visual objects for high-fidelity simulations [10]. Leveraging Unity's real-time 3D pipeline, researchers have built realistic science and technology learning tasks with strong representational fidelity and responsive interaction. This platform choice reflects Unity's strengths in rapid prototyping, asset ecosystem integration, and performance tuning for immersive education. Complementing this, Unreal Engine has emerged as the second most used platform, particularly valued for its cinematic visual quality and robust tools for complex simulations [7].

## 1.1. Immersive virtual reality technology in education

### 1.1.1. Relevant learning theories

Theories commonly applied to learning with visualizations also underpin immersive VR-based learning. Euler et al. highlight several relevant frameworks: cognitive load theory, distributed cognition, the cognitive theory of multimedia learning, Ainsworth's theory of multiple representations, grounded and embodied cognition, and social semiotics [11]. Framing VR activities through these lenses helps explain how design choices shape attention, representation, embodiment, and meaning-making during immersive learning tasks. Cognitive load theory posits that working memory is sharply limited in both capacity and duration, so excessive total load can impede learning [12]. Within this framework, cognitive load comprises three additive components: intrinsic load, arising from the task's inherent complexity; extraneous load, imposed by suboptimal presentation and other nonessential elements; and germane load, representing the resources devoted to constructing and refining schemas relevant to the target knowledge. Effective learning requires fostering germane load, minimizing extraneous load, and managing intrinsic load (for example, by segmenting information) to fit learners' prior knowledge [13]. In practice, Mayer's cognitive theory of multimedia learning operationalizes these goals with design principles; for IVR contexts, the multimedia, spatial contiguity, temporal contiguity, and segmenting principles are especially pertinent [14]. Furthermore, Wei et al. demonstrate that mental effort in virtual reality can be objectively quantified using physiological indicators such as heart rate variability and skin conductance. They further argue that the dynamic detection of cognitive load could enable the level of presence in the virtual environment to be adjusted in order to support optimal learning outcome [15].

The multimedia principle states that people learn better from a combination of words and pictures than from words alone. In IVR, this means designing experiences that blend text or narration with images, 3D objects, animations, and video, such as virtual worlds with embedded animations or pairing concise explanatory text with interactive 3D visuals.

The principle of spatial contiguity states that people learn better when related words and images are placed close together rather than far apart. In IVR, this can be implemented by positioning labels or brief explanations directly on or next to the relevant animated or physical components – for example, overlaying short text descriptions on specific car engine parts in a virtual world. Similarly, during a virtual science experiment, live measurement readouts can be shown within the learner's field of view in a headset so that data appears adjacent to the observed phenomenon. Research by Hodgkiss et al. (2025) suggests that spatial skills are more crucial for grasping physics concepts than for memorizing facts. This highlights the value of educational interventions that target intrinsic-dynamic spatial skills—specifically, a student's ability to mentally rotate, manipulate, and transform objects [16]. The principle of temporal contiguity states that people learn better when corresponding words and images are presented at the same time rather than one after the other, and the segmenting principle holds that learning improves when a multimedia lesson is divided into self-paced chunks. In IVR, both principles are straightforward to apply: interfaces can display brief text alongside relevant visuals, and experiences can be organized into a sequence of small, conceptually related modules that learners advance through at their own pace.

Although Cognitive Load Theory and Mayer's Cognitive Theory of Multimedia Learning provide a useful starting point for interpreting learning in immersive virtual reality, their applicability to IVR remains debated because these frameworks were largely developed in more controlled, non-immersive instructional settings [17,18]. Evidence from immersive science learning shows that IVR can increase learners' motivation, interest, and engagement without necessarily producing superior conceptual learning outcomes compared with well-designed non-immersive formats [17]. Furthermore, evidence for the immersion principle in multimedia learning suggests that immersive virtual environments can support learning under appropriate instructional conditions [19]. Furthermore, research indicates that immersive VR may introduce extra cognitive demands, including higher cognitive load, which can negatively shape long-term retention [20]. Recent theoretical work therefore argues for broader models of immersive learning, such as the Cognitive Affective Model of Immersive Learning, which incorporates not only cognitive load but also interest, embodiment, motivation, self-efficacy, and self-regulation [18]. Taken together, these studies suggest that CLT and multimedia learning theory should be treated as useful but partial interpretive lenses for IVR, complemented by alternative perspectives that better reflect the distinctive cognitive, affective, and embodied characteristics of immersive learning environments.

Other theories highly pertinent to immersive VR-based learning include embodied and grounded cognition, as well as distributed cognition. In embodied and grounded views, "cognitive activity is grounded in sensory–motor processes and situated in specific contexts and situations," emphasizing that perception and action within real settings are integral to thinking [21]. As already mentioned, IVR typically includes sensors that track the learner's body movement and transfer it directly into the virtual environment, which is central to embodied learning. For instance, IVR enables learners to alter the distance between two simulated bodies by directly manipulating their virtual positions.

From the perspective of distributed cognition, cognitive work is spread across internal human processes, external cognitive artifacts, and coordinated groups – so understanding and performance emerge from the interplay of people, tools, and environments rather than from an isolated mind alone [22]. External and internal representations work together during cognition, so well-designed external representations – such as IVR visualizations situated on the task context – can offload working memory and help prevent overload. For example, in the context of learning about electromagnetism, IVR-based 3D visualizations clarify the geometry of **I**, **B**, and **F**, making the right-hand rule – used to determine the Ampère force – much easier to apply. This aligns with distributed cognition accounts in which external visualizations act as “external memory,” reducing the need to mentally transform or search for information. Additionally, Patrick B. Kohl and Noah D. Finkelstein reported that representational competence is deeply tied to the specific task environment, emphasizing the need for educators to carefully design instructional activities that move beyond single-format problem solving to foster true cross-representational fluency [23].

IVR supports learning by presenting multiple, mutually reinforcing representations in the corresponding virtual-world context. Ainsworth [24] offers a systematic overview of how external representations can facilitate learning, identifying three key functions of multiple representations:

- Combining representations that complement each another (e.g., using an image to visualize the diffraction fringes’ form and a graph to present the corresponding intensity profile).
- Constraining interpretation (e.g., pairing a motion video with a real-time graph so the video anchors meaning and discourages interpreting the graph as a literal picture of the motion).
- Helping learners build relational links (e.g., mapping features of the x–t graph onto the $v_x$–t graph to connect position and velocity).

It is important to note that effective use of multiple representations requires careful design; otherwise, the combined demands can result in cognitive overload and impede learning.

Finally, because IVR often elicits rich classroom discourse, a social semiotic lens [25] is highly pertinent: researchers can examine how meaning-making resources – such as talk, gesture, body orientation, and disciplinary representations – are orchestrated in IVR learning situations to produce and negotiate meaning, for example, how overlays, labels, and shared spatial references coordinate discussion and interpretation.

#### 1.1.2. **Previous literature reviews**

Table 1 synthesizes leading reviews of IVR in education over the past decade, noting that comparatively few focus specifically on STEM. Reported sample sizes range from 29 studies in Lui et al.’s [26] science education review to 219 in Won et al.’s [27] broader education review. The most recent systematic review centered on STEM applications was published in 2024 by Acevedo et al. [7], encompassing 30 studies.

Table 1. List of past IVR education review studies.

| Authors | **N[a]** | **Research focus** |
|---|---|---|
| Matovu et al. (2023) [3] | 64 SCE | Rationales for adopting IVR in science education, identified and incorporated theories in the design, implementation, and evaluation of IVR learning, immersive design |
| | | features, evaluation of IVR learning activities, achieved learning outcomes, learners' characteristics |
| Acevedo et al. (2024) [7] | 30 SE | Database frequency, Papers per country, Papers per year, Development features, Methodological approaches for design and evaluation, Embodiment, Immersion, Type of learning, STEM topics, Learning outcomes, User experience |
| Lui et al. (2023) [26] | 29 SCE | Design of IVR lessons to optimize learning outcomes (cognitive load, learners characteristics, attitudes, knowledge retention, embodied cognition) |
| Pellas *et al*. (2021) [28] | 46 GE | Learning performance, Research methods, Instructional design, Benefits and challenges |
| Won et al. [27] | 219 GE | Technological and pedagogical features, Relation of implemented features to learning tasks, Design approaches |
| Marougkas et al. [29] | 69 GE | Benefits, Limitations, Development features, Personalization techniques, Gamification techniques, Apparatus characteristics |

[a] N = number of reviewed articles, GE = general education, SE = STEM education, SCE = science education.

Matovu et al. (2023) found that IVR was commonly selected by educators to make abstract concepts more accessible and to enhance learners' experiences. Most applications foregrounded sensory and actional components, while narrative and social features were comparatively rare. Learning theories appeared to exert limited influence on design, implementation, and evaluation. Participants generally described their IVR experiences as engaging and motivating; however, learning gains were inconsistent, and no particular immersive feature reliably produced better outcomes. Aligning immersive design features with the rationale for adopting IVR and with the chosen evaluation methods may yield more productive inquiries into IVR's educational benefits for improving science teaching and learning [3].

Acevedo et al. (2024) identified several advantages and disadvantages reported in user experience studies in the higher education context. They are proposing a conceptual framework to categorize the design and features of the IVR tools discussed in the papers, based on their levels of embodiment, immersion, and the type of learning they facilitate. It has been found that IVR can provide an intuitive and practical immersive learning experience and multiple results on motivation, engagement, usability, and learning performance are reported. Acevedo et al. (2024) recommend that designers and developers explore developing customized IVR experiences tailored for STEM learning, especially for invisible and complex subjects where 3D visualization could benefit students [7].

Lui et al. (2023) found that IVR learning outcomes improve when learners are given substantial control over their experiences, cognitive load is optimized, learner characteristics inform application design, and reflective activities are added before or after IVR use. Details of lesson designs were examined, and six design principles are proposed to support the transition to IVR lessons and enhance both application and lesson design, offering pedagogical guidance for future educators [26].

Pellas et al. (2021) review synthesizes studies reporting qualitative and quantitative evidence on current IVR-supported practices, focusing on student outcomes and performance, as well as the benefits and challenges of visual features and design elements across mobile and desktop implementations in diverse subjects. Most studies evaluated the suitability or effectiveness of instructional design processes using IVR applications, disseminating findings on user experience, usability, outcomes, and performance. The review particularly contributes to the existing literature by analyzing how instructional strategies and techniques can enhance learning with a wide range of IVR tools and by offering recommendations to guide effective instructional design across contexts for a more accurate, up-to-date picture of the field [28].

Won et al. (2023) investigated contemporary designs for learning with immersive VR by identifying nine recurrent design features. Across the literature, technological features that cultivate physical presence were more frequently implemented and examined than pedagogical features targeting engagement. Using k-means clustering, the authors delineated five approaches that vary in interactivity and task openness, ranging from passive world-watching to responding to personalized prompts. This variation reflects competing practical and educational priorities, including accessibility, interactivity, and engagement. The review highlights this diversity and recommends that future empirical work and systematic reviews attend to task design and learning experiences alongside topics and learner demographics [27].

Marougkas et al. (2024) particularly focused on personalization strategies used in classroom applications of immersive VR with respect to benefits, limitations, and development characteristics. This review is distinguished by its detailed analysis of personalization and gamification techniques within IVR, and by the observation that prior applications insufficiently leveraged opportunities for adaptive learning, indicating a clear research gap. The paper provides recommendations for future work and serves as a resource for academics, students, and professionals implementing IVR in education [29].

We can conclude that although earlier IVR review articles also included physics education studies, to our best knowledge, they did not provide topic-level overviews of immersive VR learning activities in physics. The present review contributes by detailing both pedagogical and technical aspects of IVR use for individual physics topics. In our opinion, providing an overview of IVR based activities at the level of specific physics topics, is critical for the popularization of IVR among physics educators. Such an overview is intended to encourage teachers to adopt IVR and to facilitate effective classroom implementation. In addition, our review synthesizes methodological practices in prior IVR-physics studies, which may help identify research gaps and offer guidance for strengthening IVR research methodology.

## 1.2. Study purpose and research questions

The aim of this review is to thoroughly examine the opportunities and challenges of using immersive VR (IVR) for teaching and learning physics topics. Included systems were HMD-based immersive environments, including interactive HMD applications and 360° video experiences delivered through HMDs.

We stated the following research questions:

**RQ1:** In what ways are immersive virtual reality technologies (IVR) utilized to support learning across various physics topics?
*Significance: Addressing this research question may allow us to clearly showcase the unique characteristics of IVR technology that support learning across different physics topics. By summarizing IVR-based learning activities within specific subject areas, we hope to inspire physics teachers to incorporate IVR into their teaching practices.*

**RQ2:** What are the potential advantages and disadvantages of utilizing IVR in the teaching about physics topics?
*Significance: By addressing this research question, we can develop evidence-based conclusions regarding the potential advantages and disadvantages of using IVR in physics education. This will allow us to assess the consistency of previous research findings and deepen our understanding of IVR's impact across different learning domains.*

## 2. Methods

To effectively address our research questions, we first needed to identify a selection of relevant peer-reviewed journal articles focused on IVR-based teaching and learning about physics topics. We followed the PRISMA (Preferred Reporting Items for Systematic Reviews and Meta-Analyses) [30] guidelines for selecting articles. An overview of our article selection process is shown in Figure 1.

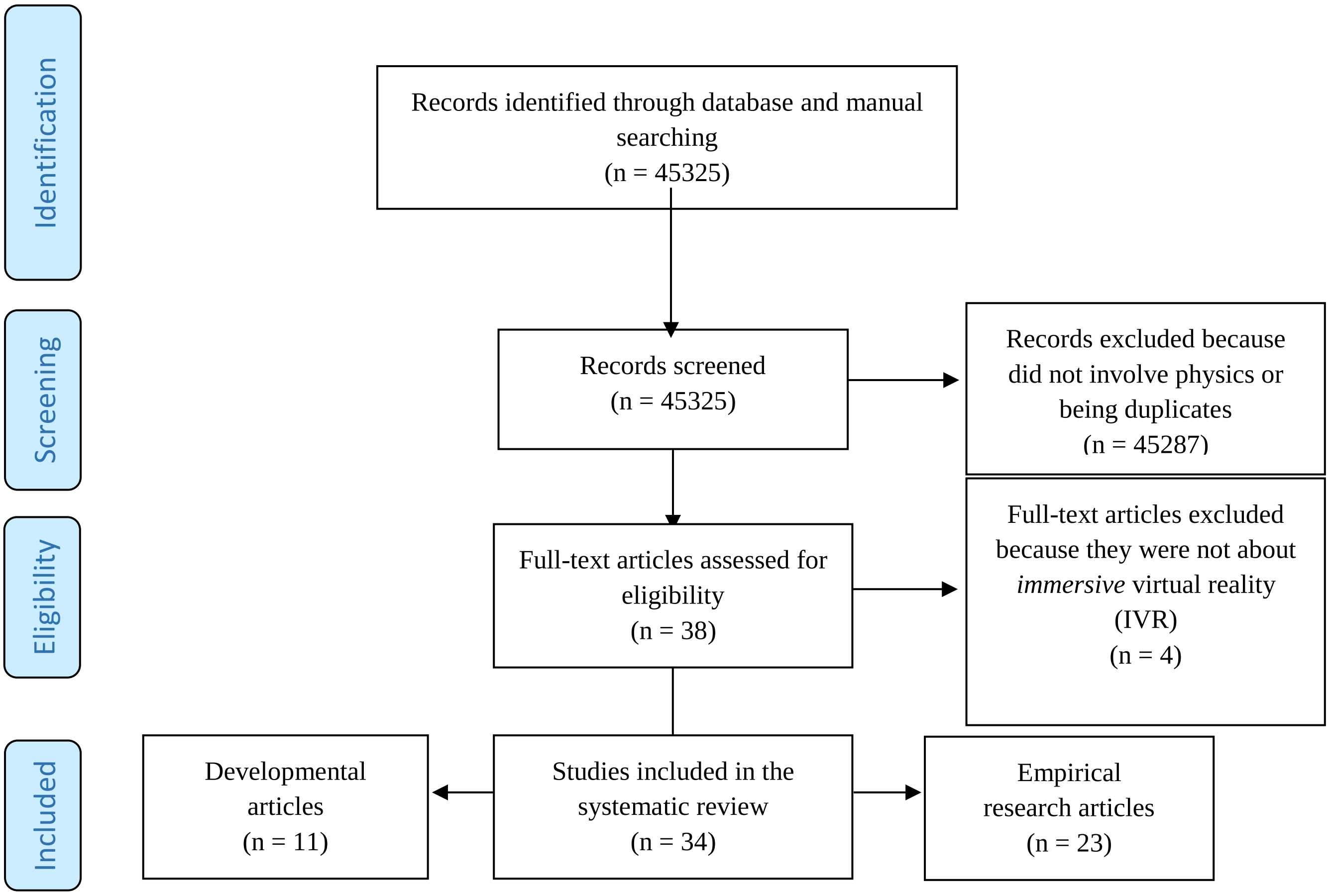

Figure 1. Article selection process based on Preferred Reporting Items for Systematic Reviews and Meta-Analyses (PRISMA) methodology [30].

The initial search was run in Scopus and ERIC on January 1, 2024 to capture studies published from January 1, 2018 to January 1, 2024; an update search was run on June 1, 2025 to capture studies published from January 1, 2024 to June 1, 2025. Full database-specific search syntax, database filters, search dates, and resulting numbers of records are provided in Appendix A. Articles were deemed relevant if they met the eligibility criteria listed in Table 2.

Table 2. Inclusion criteria.

| **Inclusion criteria** |
|---|
| Published between January 1$^{st}$, 2018 and June 1$^{st}$, 2025. |
| Peer reviewed journal article. |
| Related to using IVR in teaching about physics topics. |
| Written in English. |
| Original non-review article. |

Searches were deliberately broad. In both Scopus and ERIC, the only substantive search phrase was “virtual reality”, searched in title, abstract, and keyword/subject fields where supported by the database. The terms “physics” and names of specific physics topics were not included in the search strings. This decision was made to avoid excluding studies that addressed physics-relevant content under more specific labels such as astronomy, relativity, electromagnetism, particle physics, or laboratory simulation. Discipline-specific filtering was therefore conducted manually in two stages: first by screening titles, abstracts, and keywords, and later by screening full texts against the inclusion criteria. Searches were restricted to peer-reviewed, English-language articles published on or after January 1, 2018. This cutoff reflects the fast pace of educational technology and the aim to align pedagogy with recent advances. The year 2018 was chosen as the starting point because higher-resolution IVR head-mounted displays began entering the market at that time. Review articles were excluded to focus on original studies contributing new teaching ideas for IVR-based physics learning. To ensure a consistent level of technological immersion across the analyzed literature, “immersive VR” (IVR) in this review is strictly operationalized as systems utilizing Head-Mounted Displays (HMDs) that visually isolate the user from the physical environment. Consequently, borderline and semi-immersive systems—such as desktop-based VR, screen-based simulations, or projection-based CAVE systems that do not utilize HMDs—were explicitly excluded from the analysis.

Applying the above-mentioned criteria to titles, abstracts, and keywords returned 1,777 records in ERIC and 43,544 in Scopus. In addition, we manually found 4 relevant articles from past reviews. Records were then screened to retain only IVR studies about learning physics, where “physics” was defined as the concepts and procedures used by physicists to study natural phenomena, consistent with physics textbook content across different educational levels. Each time we recognized that the title or abstract or keywords thematize learning about some physics concept or procedure, the corresponding article has been retained.

Screening 45,325 records by title, keywords, and abstract yielded 43 potentially relevant IVR studies on physics learning. After removing 5 duplicates, 38 unique articles proceeded to full screening against the inclusion criteria and checking for basic reporting adequacy. During full-text assessment, 4 of the 38 records were excluded because they were not related to immersive virtual reality (IVR). Thus, a final set of 34 articles for the systematic review of IVR-based physics learning was obtained. Out of these 34 articles, 23 were empirical. For the purposes of this review, empirical studies were defined as those that involved the systematic collection of data concerning the experience of learning or teaching physics, whereas developmental studies were defined as those that did not include such data collection. Both empirical research and developmental studies were considered eligible as potential sources of ideas for IVR-based physics learning, i.e. for answering research question 1. On the other hand, for making conclusions about the effects of IVR-based learning, only the 23 empirical articles were considered relevant

To analyze the effects of IVR-based learning, we conducted a qualitative thematic analysis of the findings sections for all selected articles. The sentence was established as the primary unit of analysis. We employed a hybrid coding approach, combining a broad organizing framework with inductive, data-driven code generation. During the initial review of the texts, three overarching domains emerged to categorize the advantages and disadvantages of IVR: affective, cognitive, and practical. These broad domains served as the foundational structure for our analysis. As we systematically analyzed the texts, we applied a structured coding convention to capture the valence (advantage or disadvantage), the overarching domain, and the specific emergent concept identified in the sentence (e.g., advantage_affective_flow, advantage_cognitive_relevant_load, disadvantage_practicality_health). To facilitate theme development, the coded excerpts were extracted and charted into six distinct analytical matrices based on their domain and valence (e.g., Affective Advantages, Affective Disadvantages, Cognitive Advantages, etc.). Within these domain-specific documents, we iteratively reviewed and synthesized the grouped excerpts to construct and define the final overarching themes regarding the impact of IVR.

The whole initial review was done by the first author alone. For reliability, a second author independently screened a random 25% subsample of the initially identified records against the inclusion criteria; inter-rater agreement was 92.9% before discussion, after which consensus was reached. Both authors have more than 15 years of physics teaching experience.

# 3. Findings and discussion

In this section, we outline and analyze the results corresponding to each of the previously stated research questions.

## 3.1. In what ways are immersive virtual reality technologies (IVR) utilized to support learning across various physics topics?

Table 3 highlights key concepts related to IVR-based learning and teaching physics, and also offers a summary of the research designs that were implemented.

Table 3. An initial overview of the literature reviewed, highlighting IVR-based learning and the research designs used; all 34 articles are included in this summary.

| Authors | Topics | Research design | IVR-based learning |
|---|---|---|---|
| Acevedo, Magana, Walsh et al. https://doi.org/10.1016/j.cexr.2024.100075 [31] | Coulomb's law and electric field lines | Pre-test-post-test control random group qualitative and quantitative experimental design was used. Data was collected through the interviews, screen capture, SUS questionnaire, User perception questionnaire, Conceptual and Representational questions with open ended and multiple-choice questions. | Head mounted display (HMD) with haptic controllers was used to run immersive VR simulation to learn about Coulomb's law and electric field lines. Undergraduate university students grab and move virtual charges with their hands to change configurations, linking bodily action to changes in forces, distances, and equipotentials, which strengthens memory traces and supports agency in learning. |
| Alka et al. https://doi.org/10.26803/ijlter.23.11.13 [32] | Optics: the Michelson-Morley experiment | Developmental article | This immersive virtual reality environment replicates the experimental setup for Michelson-Morley experiment. Students can navigate the virtual lab, select components, and manipulate controls to align the interferometer and optimize fringes. This procedural engagement mimics authentic experimental work, helping students develop experimental reasoning even though they are in a simulated environment. Instead of requiring a delicate interferometer and vibration isolation, students can run the simulation on a gaming laptop with an IVR headset, removing many logistical barriers. |
| Bancong and Nanda https://doi.org/10.1088/1361-6552/adb08a [33] | Photoelectric effect | Developmental article | Immersive VR environment simulates photoelectric effect experiment that replaces expensive and often unavailable photoelectric setups with a virtual laboratory, so students can perform the experiment even in schools without real equipment. Students can manipulate light frequency and intensity to observe electron emission and analyze relationships between photon energy, work function, and electron behavior. They can explore the responses of 25 different metals, study the kinetic energy of emitted electrons, and investigate the effects of light intensity on photocurrent. The simulation provides interactive features like tutorials and quizzes and let students record numerical results in simulation and analyse relationships between variables. |
| Bedregal-Alpaca et al. [34] | Celestial bodies and planetary systems | Researchers used questionnaires inside the IVR application and administered survey on the perception of users as to the usefulness of the application and the satisfaction of its use. | Students (first degree of secondary education) used Google Cardboard with Android mobile device to run IVR system that helps by immersing students in astronomy through an interstellar spacecraft journey with interactive modules and mini-games. It enhances motivation and learning by offering quizzes, immediate feedback, rewards, and a virtual guide. Accessible via mobile IVR devices, it visually explains key concepts like lunar |
| | | | phases and galaxies, making learning engaging and effective. This combination of IVR and gamification can increase student participation and interest in astronomy. |
| Burt and Pendrill https://doi.org/10.1088/1361-6552/ab9872 [35] | Kinematics | Developmental article | The IVR representation simulates a Drop Tower Ride from a first-person perspective, i.e. users experience the ride virtually, seeing what a rider would see during free fall and bounces. Also, virtual ride offers the possibility to overlay force and acceleration vectors. This simulation provides students a visual and immersive experience of the motion inside the ride, helping them connect abstract physics concepts (such as acceleration and force) with a more intuitive, bodily experience. |
| Canright and White Brahmia https://doi.org/10.1103/PhysRevPhysEducRes.20.010146 [36] | Coulomb's law (fictitious laws from another universe) | Researchers utilized pre-post measurements to collect data for epistemology about experimental physics, self-efficacy and flow. Quantitative and qualitative data analyses were performed. | Each group of undergraduate students shared one IVR headset with its display mirrored onto a lab computer in the IVR system that helps by immersing students in a virtual lab where they explore and model new, fictional physics laws. It encourages students to develop their own models through experimentation rather than simply confirming known theories. The system provides realistic tools for measurement and manipulation, promotes collaboration, and increases students' confidence and engagement. By simulating research processes and allowing the study of impossible scenarios, the IVR system fosters expert-like thinking and deeper learning about the nature of experimentation in physics. This approach leads to more authentic, engaging, and meaningful lab experiences than traditional methods. |
| Chen and Syu https://doi.org/10.1111/bjet.13436 [37] | Astronomy (star constellations) | Two group quasi-experiment was conducted. The measuring tools used in this study included the pretest, posttest and surveys of self-efficacy, motivation, perception of immersion, learning effectiveness and interview with selected students. | Students (fifth-grade, 10-11 years) used headsets to observe a 360° starry sky with realistic stars, links, and constellation diagrams from astronomy database that respond to head movement, creating a sense of presence and self-location in an authentic-looking night sky. Students pilot a spacecraft to move through the environment while carrying out learning missions. Interactive environment wrapped in a guided role-playing game that structures tasks, gives immediate feedback, and keeps students focused and motivated, which in turn increases immersion, self-efficacy, and extrinsic motivation. |
| De Angelis et al. https://doi.org/10.1088/1361-6552/ac93dd [38] | Particle physics, Quantum physics, electrodynamics | Developmental article | Immersive VR application simulates a visit to a real particle physics research laboratory where students can explore a particle detector. IVR application simulates realistic particle interactions and students can pause, rewind, or fast-forward the motion of particles to study how they interact with each other. The authors highlight that the IVR visit have potential to foster motivation and interest in physics/STEM careers by letting students "see through the researcher's eyes" and |
| | | | share their emotions and experiences in the VR lab context. |
| de Moraes Rossetto et al. https://doi.org/10.1002/cae.22671 [39] | Solar system | Researchers used pretest/posttest quantitative and qualitative research methodology to verify students' experiences and knowledge. | Students (sixth-year middle school) used the IVR application with smartphone and attached low cost Google Cardboard to see visualization of the planets of Solar system and information about them with increasing sense of interactivity, presence and attention. Students see each planet as a 3D model accompanied by an information panel that presents attributes such as average distance from the Sun, temperature, potential for life, rotation period, translation period, diameter, planet type, and number of moons. Navigation arrows within the IVR scene allow students to move systematically through all planets of the solar system. |
| J. Franklin and Ryder https://doi.org/10.1119/1.5080224 [40] | Electromagnetic field | Developmental article | Students use HTC Vive virtual reality system to utilize simulation that helps by providing immersive, three-dimensional visualizations of electromagnetic fields, both static and dynamic, allowing users to build intuition and understanding of these complex concepts. The system enables users to place charged spheres in a 3D grid dynamically, and instantly see the resulting electric field vectors in space, demonstrating superposition and configurations like line charges and charged sheets. Additionally, the IVR environment allows users to create paths along which a charged particle moves at adjustable speeds, including relativistic speeds, and visualizes the resulting electric and magnetic fields as heat maps and vector fields. |
| Georgiou et al. https://doi.org/10.1007/s11423-021-10055-y [41] | Special theory of relativity (length contraction, time dilatation in space) | Researchers adopted pretest/posttest mixed-methods research design. The data were collected using conceptual test, adapted Constructivist Multimedia Learning Environment Survey, and subset of students participated in a semi-structured individual interview. Quantitative and qualitative data analytical techniques were used. | High school students (10th–11th graders) run immersive VR space simulation and use head-mounted display with body tracking and controllers connected with PC to observe effects of special theory of relativity when traveling by spaceship. Simulation enables students to engage with interactive, realistic graphics and representations, making invisible or unobservable relativistic effects visible and intuitive (e.g., traveling in a spaceship at relativistic speeds through the solar system and observing the Lorentz contraction across the Milky Way). IVR is embedded within an inquiry model—students accomplish missions (such as the "twin paradox"), manipulate key variables (speed, destination, reference frames), verify hypotheses using VR data, and make evidence-based conclusions like scientists, all scaffolded by collaborative data collection and reflection. |
| Huang et al. [42] https://10.1089/cyber.2018.0150 | Astronomy (Solar system) | Researchers tested their hypothesis using two randomly assigned student groups. Pretest and posttest questionnaire about psychological and | College students (average age 20,5 years) interacted with a mobile app called "Solar System Space Museum" on a smartphone paired with a VR headset to explore three-dimensional models of the solar system. |
| | | cognitive feelings of spatial presence and science knowledge were used. | The app provided automated commentary on astronomy and detailed facts about the planets. During the five-minute experience, participants held in the hands the VR headset and observed the digital content against a white background. |
| Kersting et al. https://doi.org/10.1080/21548455.2020.1857458 [43] | Celestial bodies and astronomical phenomena | Researchers used case study qualitative design where they used survey, observation notes, video recordings, VR screen captures and focus group semi- structured interviews with outreach and science professionals. | Science festival visitors (mostly families with children) are using wireless head-mounted IVR-headsets to visualize otherwise abstract, difficult-to-grasp phenomena—such as black holes and the scale of cosmic objects—in visually striking and interactive ways. Science professionals cycle through the scenes and guide visitors through the tour. This IVR simulation places visitors directly into a virtual universe, allowing them to move around and interact with astronomical phenomena, resulting in an increased engagement and strong sense of presence and immersion while headsets devices connected to shared computers are enabling multiple users to explore at once, which amplifies the collaboration and feeling of "being there", and lets visitors experience phenomena such as floating or flying through space. |
| Korlat et al. https://doi.org/10.3389/fpsyg.2024.1284597 [44] | Electricity, electrical voltage, electrical currency, electromagnetic spectrum, forces, climate, energy, temperature, magnetic coil and radioactivity | The researchers employed a quasi-experimental two-group pretest/posttest research design. They used questionnaires to assess students' interest, self-efficacy, self-concept and entity implicit theories regarding physics. | Secondary school students used game-based low-cost IVR system with mobile phone application and Google cardboard to learn about different physics topics. A robot character guides student, explains concepts via recorded audio, and acts as quiz master, giving a narrative frame and instructions inside the virtual lab spaces. Students can actively experiment with otherwise hard-to-realize and abstract phenomena, which can support motivation and self-beliefs in physics. |
| Kortemeyer https://doi.org/10.1119/5.0067963 [45] | Kinematics | Developmental article | Using IVR hand controller students can grab, move, and throw virtual objects (ball) and with HMD observe the velocity and acceleration vectors of ball in real time which can help realizing that during free fall the acceleration vector always points downward, even when the velocity is upward along the trajectory. Furthermore, students cab use IVR system to demonstrate circular motion concepts that are otherwise hard to visualize. Additionally, The IVR view can be mirrored as a clipped-out section on a regular computer screen or projector, so while one student is in VR, the rest of the class can watch what happens and discuss or give directions which turns an individual immersive experience into a shared classroom resource. |
| Kwon https://doi.org/10.1007/s10055-018-0364-1 [46] | Astronomy | The researchers conducted quasi-experimental two group research where groups were selected ensuring an even distribution of students with high and low academic achievement. Utilized 5-point Likert scale survey | Students (grade four, 11 years old) use HTC VIVE HMD with multi-dimensional controller that allow them to actually walk in a designated space and naturally touch or grab virtual objects. Utilized Moon Exploration Learning Game is an adventure |
| | | consisted of variables for measuring the recognition of the directness of experience (presence, vividness, tactile interactivity, locomotive interactivity, and simulator sickness), the requisite of the learning effect (flow), and the variable for measuring the learning steps (experientiality). Additionally, to measure the learning effect an academic achievement essay question sheet was used. | game with a story, role and quest that provides realistic gaming experience where students can learn about concepts of celestial bodies, space and gravity. |
| Lee et al. https://doi.org/10.3390/app11198932 [47] | Sun's influence on Earth and near-Earth Space | Survey research was conducted using Likert-type questionnaire for data collection. Data were analyzed quantitatively and qualitatively. | The IVR system in the article helps by providing an immersive virtual reality (VR)-based streaming distance education platform specifically designed for the Solar Dynamics Observatory (SDO) data. It enables multiple remote users, including experts (scientists) and non-experts (students), to share, experience, and interact with real-time solar observation images in a virtual space using head-mounted displays (HMDs) and controllers for navigation and participation. |
| Madden et al. https://doi.org/10.1371/journal.pone.0229788 [48] | Astronomy (Moon phases) | Researchers employed a three-group random pretest/posttest design. The pretest evaluated students' understanding of the Moon phases, while the posttest assessed their knowledge of the Moon phases and included surveys on demographics, attitudes, video game and VR experiences, academic major, and environmental views. After completing the tests, students were invited to experience the other two conditions and were asked to fill out a brief survey about their experience. Additionally, a delayed posttest was administered four months after the experiment. | Undergraduate students wore the Oculus Rift headset and used handheld controllers that tracked their movements, enabling them to interact with the virtual environment. The simulation represented the Sun-Earth-Moon system, allowing students to adjust time to observe changes in the Moon's phases, physically "grab" and move the Moon along its orbit, and switch between various perspectives. Additionally, the system featured a "virtual clipboard" attached to the user's hand, which displayed instructions and questions. |
| Matys et al. https://doi.org/10.3390/photonics12050436 [49] | Plasma physics | Developmental article | In immersive VR mode of the application students can use HMD system to explore a variety of high intensity laser-matter interactions such as ion and electron acceleration, gamma-flash generation, electron-positron pair production or attosecond and spiral pulse generation. They can zoom, rotate and move within the simulation space using body movement or controllers creating immersive, interactive 3D view of laser–plasma simulations, making it easier to see particle dynamics, fields and structures in space and time than with static 2D plots or videos. |
| Mufit et al. https://doi.org/10.36681/tused.2024.020 [50] | X-rays, black body, Compton effect, photoelectric effect | Developmental article with design research through the stages of needs analysis, product design, validity test and practicality test. Qualitative and quantitative data was collected using questionnaires. | First year physics students used IVR headset to run application and learn about black body, photoelectric effect, Compton effect and x-ray production arranged as in a real lab, with all tools modeled in 3D and placed on benches and cabinets. Each experiment is scaffolded with embedded video tutorials (on using IVR), on-screen equations and formulas in virtual books, and a virtual calculator and computer for data processing and writing conclusions, so |
| | | | the entire inquiry cycle happens inside IVR. Students manipulate virtual apparatus, move around, and operate instruments, which increases a sense of presence and active engagement. |
| Mufit et al. http://dx.doi.org/10.62527/joiv.9.1.2244 [51] | X-rays, black body, Compton effect, photoelectric effect | A research design in multiple phases was conducted. Researchers first randomly selected students and conducted needs analysis by questionnaires that consisted of Linkert scale and open-ended questions. After VR implementation performance evaluation was carried out using validity instruments by experts and practicality instruments during the practicality test by students. | Physics education students (18-19 years) used IVR headsets to run digital simulations and interact with virtual environment objects in the X-rays, black body, Compton effect, photoelectric effect experiments. The IVR system helps by providing an interactive, immersive quantum physics lab where students can perform key experiments virtually. It makes abstract phenomena visible, replicates real lab workflows, allows safe and repeatable practice, engages users through embodied interaction, and improves accessibility and learning effectiveness. This supports deeper understanding and practical skills without needing expensive or hazardous real equipment. |
| Planey et al. https://doi.org/10.1080/10494820.2023.2239290 [52] | Astronomy | Researchers collected audio, video and software interactions of student groups that were working on tasks. Qualitative techniques were used to analyze the collected data. | During the learning about the astronomy (ie. find the position of a space capsule) undergraduate students were able to view the night sky at different dates, times, and locations using the VR headset. Student repeatedly finds stars, reports their altitudes, and these values are used by other students to propose, confirm, reject, or modify ideas about capsule position. One student with IVR headset acts as a data gatherer while tablet users manage the data worksheet. IVR system helps by making star data easy to obtain and share, by enabling precise joint visual reference across devices, and by structuring complementary roles in problem solving. |
| Porter et al. https://doi.org/10.1103/PhysRevPhysEducRes.16.020119 [53] | Electrostatics | The researchers employed a two-group pretest–posttest randomized research design. Questionnaires were used to assess students' knowledge of electrostatics. The pretest consisted of 13 multiple-choice questions, while the posttest consisted of 11 multiple-choice questions. Upon completing the electrostatics-related questions, students were asked to evaluate the VR intervention using a 5-point Likert scale, after which they answered demographic questions. Additionally, three questions were integrated into the VR simulation to monitor students' engagement with the material. | Second-year college students used smartphones inserted into simple headsets that delivered stereoscopic VR images and responded to head movements. The app presented several electrostatics scenes (dipole, charged rod, charged plate) and allowed students to explore electric field vectors from multiple viewpoints. |
| Šiđanin et al. https://doi.org/10.1088/1361-6404/ab9c90 [54] | Nuclear physics | Developmental article | Using the IVR headset and a pair of controllers, students carry out a nuclear physics experiment to determine the mass of the deuteron. They simulate placing a neutron source and water container near a detector, record gamma spectra, and observe how different setups affect the results. IVR system familiarizes students with nuclear physics phenomena without needing actual expensive or restricted |
| | | | equipment, enables hands-on manipulation of virtual equipment, visualizes processes at atomic and particle scales and speeds up long measurement times for practical training while maintaining real data characteristics. |
| Stella et al. https://doi.org/10.1007/s11042-022-13410-0 [55] | Solid state physics: crystal structures | Developmental article | With IVR head mounted display students can explore complex 3D crystal structures (view unit cells, lattice geometries, and Miller index planes) in an interactive environment. They can rotate, zoom, pan, and apply symmetries or cuts to crystal models. This IVR environment allows arbitrarily large, clutter-free crystals, quick switching among many representations, and complex operations (symmetries, cuts, epitaxial crystals) that would be impractical with physical models. The immersive headset blocks out the real surroundings and presents a large field of view with depth cues, increase focus, and effective convey 3D shape. |
| Tai and Hong https://doi.org/10.1007/s10639-024-12860-3 [56] | Astronomy (stars and constellations) | The researchers implemented a quasi-experimental pretest/posttest design. Participants completed a questionnaire assessing self-efficacy, learning interest, frustration, and flow experience. Quantitative data analysis was conducted. | Museum visitors using IVR system observe sky to locate stars with different levels of brightness and to draw lines between stars to form constellations. By walking, looking around, selecting stars of different size, and drawing lines in a surrounding virtual sky, visitors practice the spatial layout of constellations in a way that is intended to transfer to naked-eye stargazing. Combining 3D immersive stargazing tasks with haptic and visual scaffolds (Gollin effect), the system demonstrated effective support in learning. |
| Tito Cruz et al. https://doi.org/10.1007/s10055-023-00824-w [57] | Kinematics (projectile motion) | Mixed methodology with user-based experiments, knowledge and usability tests, engagement and presence evaluations, screen capturing, and statistical analysis were performed to obtain and analyze data. | High-school students wore headsets in a room-scale IVR game on projectile motion that let them walk naturally, aim a cannon, and manipulate variables such as angle, speed and flight time while seeing trajectory markers and vector overlays. A tutorial acclimatized them before timed game levels with narrative challenges. |
| Tsivitanidou et al. https://doi.org/10.1007/s10956-021-09924-1 [58] | Special theory of relativity (length contraction, time dilatation in space) | A pretest/posttest quantitative approach was used. The data sources included attitudes presurvey, pretest/posttest conceptual test and a post-activity questionnaire on students' perceptions of the learning experience. | Using an immersive VR spaceflight simulation with tracked headsets and simple controls, high-school students travelled at relativistic speeds and observed how distance, speed, proper time and Earth time diverged to experience time dilation and length contraction. High immersion kept focus on interpreting relativistic effects instead of controlling the interface. |
| Urhan & Akpinar https://doi.org/10.15354/sief.24.or550 [59] | Sun system and eclipses | A qualitative study used semi-structured interview form as a data collection tool. | Students (6th grade of elementary school) use virtual glasses to see virtual solar system and sensors to detect the user location and direction change in application. IVR system offers an interactive, immersive environment where students start in a virtual cabin, use telescopes in an observatory, travel through space, and actively manipulate virtual simulation using hand controllers and Leap Motion technology. This realism allows students to "walk in space," observe phenomena like eclipses from different |
| | | | perspectives, and classify inner and outer planets by physically moving objects—offering “learning by doing”. The technical setup allowed simultaneous viewing on classroom monitors, increasing both individual immersion and collective engagement. |
| Van Acoleyen and Van Doorsselaere https://doi.org/10.1119/10.0001803 [60] | Special theory of relativity (length contraction, time dilatation, Doppler effect, light aberration) | Developmental article | Users utilize an IVR headset system specifically designed to provide an immersive and tangible understanding of special relativity by allowing users to experience a virtual boat trip in a world where the speed of light is drastically reduced. Students are placed inside a 360° video, letting them "ride" a boat through a city at relativistic speeds, thereby allowing direct and intuitive observation of special relativistic phenomena such as length contraction, time dilation, Doppler shift, and light aberration. Used in classroom and other settings, users can engage in collective and individual exploration, enabling both guided discussion and playful, curiosity-driven investigation of special relativity. |
| Villada Castillo et al. https://doi.org/10.3390/app15063405 [61] | Projectile motion | Study implemented a single- case pretest and posttest experimental design. Data was collected using questionnaire that consist of multiple-choice questions that assessed students’ theoretical understanding of given physics phenomena. | University students (second semester) use a head mounted display and hand controllers to run realistic IVR game “Parabolic basketball VR” designed as an educational tool for teaching projectile motion. The environment provides continuous visual feedback on relevant variables (e.g., angle, velocity, trajectory) and gives positive reinforcement or corrective guidance depending on player performance. Student interaction is embodied (throwing with the virtual hands in a simulated stadium) and abstract kinematic relations are visualized (independence of horizontal/vertical motion, horizontal launch, optimal 45° angle, etc.) which results in 3D learning experience. |
| Wang https://doi.org/10.1007/s10639-024-13156-2 [62] | Electricity, Thermodynamics, Mechanics | Developmental article | Participants use a head-mounted display device to go through three stages, each with different theme and experimental setup: electrical and thermal conductivity, and ductility of metals and nonmetals. They can pick up items, move them around and interact with experimental equipment that can potentially increase presence, attention, and engagement compared to traditional lectures. Additionally, students can repeat risky or equipment-intensive procedures virtually without danger or material cost. |
| Xie et al. https://doi.org/10.1109/TLT.2023.3263587 [63] | Electrical circuits | In a randomized controlled study, students were administered a pretest and posttest instruments to measure academic self-efficacy, intrinsic learning motivation, cognitive load and learner’s satisfaction as well as academic achievement. Quantitative analysis of data was used. | Each student (secondary school, 13 to 15 years) group utilizes a high-performance computer equipped with a high-definition wired head-mounted display, controllers, and laser positioning sensors. IVR lab simulates an authentic space with whiteboards, tables, chairs, and storage cabinets, giving students a realistic lab context even if a real lab or equipment is limited. Students connect simulated components—including batteries, |
| | | | switches, light bulbs, and an ammeter—during the experimental task. Furthermore, students receive either corrective feedback or explanatory feedback. The IVR system provided students an interactive, immersive virtual lab that can increase their motivation, self-efficacy, satisfaction, and test scores without adding extra cognitive load compared with video. |
| Yulkifli et al. https://doi.org/0.18178/ijiet.2025.15.6.2332 [64] | Rotational dynamics | Researchers used quasi-experimental method with two randomly selected student groups. Data collection in this study used an instrument test of creative thinking skills and self-efficacy. | First-year university students enter a virtual scene using IVR glasses with a rolling barrel on an inclined plane. They conduct virtual experiments by modifying height, slippery and roughness of inclined planes. The IVR scenario requires students to operate IVR glasses and a joystick, explore the environment, and control experiments themselves, directly increasing creative thinking and self-efficacy indicators such as confidence in using IVR features and learning effectively with IVR. |
| Zöllner et al. https://doi.org/10.1016/j.cag.2025.104258 [65] | Nuclear physics | Developmental article | Participants use a head-mounted display device to learn about abstract decay rows through different ways of interacting with the nuclide chart: teleportation with controllers, physical walking, or hand gestures. They can select nuclides and follow decay series to reach a stable state. By selecting individual nuclides, students can verify if they are part of a decay series, and receive feedback (visual, auditory and haptic) based on their selections. Embodied navigation on floor can strengthen spatial understanding of how decay series move across the chart while wall variants support an overview perspective: students stand back and use ray-casting or finger touch to select nuclides and move the chart section on the wall. |

From Table 3 we can conclude that a wide range of IVR features can be used to facilitate learning of various physics topics. By analyzing the different learning activities, we identified the following ways in which IVR may support physics learning:

1) Facilitating learning by enabling embodied enactment, where body tracking and hand controllers map a learner's motor actions directly to physics concepts, thus strengthening understanding through sensations of presence and embodiment e.g. [31]
2) Facilitating learning by visualizing abstract phenomena and simultaneously reducing extraneous cognitive load, as complex concepts (e.g., constellations, solar systems, electric fields) can be presented in an intuitive, spatiotemporal 3D format that eliminates the need for learners to mentally translate two-dimensional representations e.g. [9, 43]
3) Facilitating learning and causal reasoning by allowing interactive variable manipulation with immediate feedback (e.g., projectile angle/velocity, surface roughness, incline height) e.g. [57, 64]
4) Facilitating learning by signaling and scaffolding (e.g., annotations, highlights, narrative structure) e.g. [35]

5) Facilitating learning by immersion into simulated, trustworthy reconstructions of risky or highly sophisticated laboratory environments (e.g., X-ray production, Compton effect), which allows for a reliable, safe, and repeatable practice that effectively complements the development of important laboratory skills e.g. [51]
6) Facilitating learning by enabling a structured collaborative orchestration that uses headset mirroring and defined group roles, allowing observing peers to actively discuss, reflect on, and critique the operator's actions in real-time e.g. [47]
7) Facilitating learning by allowing comprehensive control over time and location within a virtual environment, thereby enabling learners to investigate phenomena that are otherwise inaccessible due to spatial or temporal constraints (e.g., investigating celestial motion by changing date/time/location) e.g. [52]

Mostly, IVR applications facilitate learning by providing multiple representations (visual, auditory, and haptic). In fact, providing multiple complementary representations of some phenomena may positively impact learning (Ainsworth, 2008). Concretely, Ainsworth states that learning benefits when different representations contribute non-redundant information or support distinct cognitive processes that each add something unique to understanding physics phenomena. For example, IVR can achieve this by combining immersive spatial scenes with corresponding symbolic, graphical, textual, audio, and haptic elements (e.g., a student's movement may be tracked with a motion sensor and the corresponding movement graph/tabular data is displayed on the screen). In other words, IVR can co-present 3D experiences and linked external representations, so learners gain insights unavailable from any single representation alone.

It is important to note that learning with multiple representations can overload working memory when representations incorporate features irrelevant to the learning objectives; for instance, in gameplay-like contexts, off-screen threats or startling “seductive details” can capture attention and may foreground irrelevant surface attributes and hinder conceptual focus [57]. To support more objective evaluation of instructional effectiveness, future IVR studies should disaggregate and report the cognitive load components elicited by activities, distinguishing intrinsic, extraneous, and germane load.

When it comes to the research design, it should be noted that qualitative inquiry can yield rich insight into IVR-mediated learning processes, thereby informing more precise hypotheses for subsequent quantitative testing. In the studies reviewed, most authors triangulated multiple data-collection methods to evaluate IVR in physics instruction. Written assessments were the most common instruments, mirroring broader trends in physics education research, while observational measures were also frequent which are particularly well suited to capturing competencies associated with scientific-method practice [66].

In developmental papers, typically a new IVR application is presented which sometimes includes also a needs analysis, as well as a description of the application’s iterative design and the didactic opportunities it offers.

## 3.2. What are the potential advantages and disadvantages of utilizing IVR in the teaching about physics topics?

This section evaluates the potential advantages and disadvantages of IVR-based physics learning activities, based on the results from the 23 empirical articles included in our systematic review. A meta-analysis of IVR effects was not performed, as the included studies displayed substantial heterogeneity in research contexts and methodological designs.

### 3.2.1. Potential advantages

By surrounding learners with manipulable, spatially faithful representations, IVR can heighten presence and interest, which in turn sustain attention and effort. When paired with clear goals, feedback, and opportunities for agency, these motivational gains are more likely to lead to deeper processing and persistence. Eventually, the size of the advantage depends on instructional design – what students are asked to do, how information is signaled, and how collaboration is structured – rather than on the headset alone.

#### 1. Positive effects on affective domain of learning

The affective domain – emotions, motivations, values, attitudes, and self-beliefs – profoundly shapes how learners approach, persist in, and make meaning from study, with measurable impacts on performance and long-term course and career choices [67]. Motivation is the internal state that arouses, directs, and sustains behavior [68], whereas interest blends feeling-related attraction with value-related meaning in a subject [69]. As interest develops, engagement typically follows – focused attention, participation, and curiosity [70]. Furthermore, when challenge matches skill, goals are clear, and feedback is timely, learners can enter flow, a deeply absorbed and enjoyable state that supports high performance [71]. Two belief constructs are pivotal for durable gains: attitudes toward the subject (feelings, beliefs, values) and self-efficacy (belief in one's capability). It is also important to note that pedagogy is the lever that converts initial affect into sustained outcomes: inquiry prompts, scaffolding and fading, collaboration, role-play, and game-like goals help transform momentary excitement into lasting confidence, persistence, and achievement. Because novelty fades, tasks must remain meaningful**,** feedback actionable, and social structures supportive for motivation and learning to endure [42, 67-71].

Next, we provide an overview of IVR effects on the affective domain of learning physics. First, it should be pointed out that IVR activities help fulfil affective outcomes through their feature of presence or immersion. In fact, learners frequently report strong sense of presence and immersion, sometimes overlooking the medium itself and describing the environment as comfortable and interesting [47]. Presence draws curiosity and attention, which sustains exploration such as in astronomy-style and "spacewalk" experiences [43, 59]. Correlational work in the context of game-based IVR links immersion to flow [57]. Compared with alternatives, IVR is more successful than AR or static media in perceived immersion and interaction [39].

Earlier research also reports IVR effects on students' interest and intrinsic motivation. Across inquiry VR, serious games, and planetarium-like experiences, students consistently describe IVR as interesting, enjoyable, and fun, with increases in intrinsic motivation [37, 41, 48, 54, 59, 61, 63]. Novelty contributes initially but does not explain sustained engagement. For sustained engagement task design and sequencing prove to be very important [36].

Another important result from earlier research is that IVR activities may raise academic and physics self-efficacy and easily align with AAPT-type lab objectives [36, 63, 64]. One study suggests gender-differential trajectories in self-efficacy: male decline in control only [44]. Generally, earlier research supports the idea that heightened self-efficacy leads to an increased interest in physics learning, which induces a state of flow and thus facilitates learning progress [56]. At the same time, frustration was shown to be inversely related to interest and flow.

It has been shown that IVR technologies may improve the social-affective climate, including belonging, participation, and reduced hesitancy. Concretely, live IVR with audio or avatars, shared screens, and guest experts increases real-time participation and question-asking. At the same time, social presence feels "like being in the same room" [47]. Furthermore, shared simulation views orient groups toward co-constructing new knowledge and often trigger productive bursts of discourse. Thereby, higher task complexity diversifies collaboration roles [52]. Public-facing implementations show reciprocal engagement: visitors feel present and curious, while facilitators report satisfaction and richer conceptual conversations [43].

Finally, IVR may also positively affect attitudes toward physics and technology (ICT). After learning with IVR, many students report seeing technology as useful, contemporary, and empowering. This effect is most prominent when IVR is embedded in inquiry-rich, hands-on sequences with scaffolding [34, 41, 50, 51].

It is very important to note that certain instructional and IVR technology design features might amplify IVR's benefits for the affective domain of learning:

- **Inquiry and sequencing -** stage-based, observation-first IVR sequences deepen interest and keep attention coherent [41,50].
- **Role-play and narrative -** role playing game (RPG) approach significantly boosts engagement, total immersion, and motivation beyond content-only VR [37].
- **Serious games with familiar contexts -** for example, a sports-framed IVR ("Parabolic Basketball") channels everyday digital culture to drive attention and motivation [61].
- **Live social presence -** avatars, voice, shared screens, and expert drop-ins reduce hesitancy and heighten participation [47].
- **Shared viewpoint for groups -** a common simulation view orients teams to co-construct knowledge and engage in productive discussion [52].

- **Scaffolding and stations -** well-sequenced supports and hands-on rotations maintain motivation under time constraints [41].
- **Gamification** – when properly introduced, it increases motivation, satisfaction, and practice [34, 46].

In conclusion, when thoughtfully designed and socially orchestrated, IVR improves affective outcomes central to learning physics. IVR features such as presence and social co-presence influence immediate affective states like curiosity and attention, which facilitate the development of affective constructs such as interest, engagement, attitudes, and self-efficacy. In other words, with IVR students feel more interested, confident, engaged, and willing to participate, often entering productive flow states that support persistence and, ultimately, learning progress. The strongest implementations treat IVR not as a spectacle but as a pedagogical amplifier. IVR tools help turn hard-to-access phenomena into lived experiences while cultivating a collaborative, motivated classroom climate. In short, presence ignites interest, good design sustains it, and self-efficacy and flow carry it forward. Because novelty gives only a temporary lift, tasks must be meaningful and social structures supportive if we want motivation to endure.

## 2. Positive effects on cognitive domain of learning

Next, we provide an overview of IVR effects on the cognitive domain of learning. The first finding from our systematic literature review is that authentic inquiry and joint attention facilitate model-based reasoning. Concretely, IVR supports inquiry at students' own pace with safe hypothesis testing and immediate, shared feedback. Live teacher streaming and shared screens allow misconceptions to be addressed in the moment, improving accuracy and test gains [47]. In astronomy contexts, IVR acts as an attention hook that, combined with shared displays and talk, yields coordinated exploration [43]. Students report that some ideas would have remained unclear without the IVR experience [59]. Patterns such as a bystander directing the headset wearer ('active guiding') and then passing the headset to another person ('passing on') spread the roles and keep the group jointly focused; this effect is strongest when an expert nudges the conversation toward the physics ideas rather than just the visuals [43]. Generally, shared viewing and facilitation lower coordination costs, freeing resources for explanation, comparison, and model revision.

Another important finding is that IVR technologies facilitate visual learning and help students develop representational competence. Across domains – astronomy, electrostatics, mechanics, and relativity – IVR's lifelike, manipulable visuals help learners link abstractions to experience, remember ideas, paid more visual attention, and coordinate multiple representations [42, 43, 59]. In electrostatics, IVR cues improved diagram quality, verbal explanations, and depiction of field lines [31]. In mechanics, variable manipulation and real-time trajectory observation shift learners from rote recall to conceptual mastery, supporting hypothesis testing and iterative refinement [61]. In special relativity, immersive visualization along with inquiry clarifies axioms (invariance, constancy of *c*) and consequences (length contraction, time dilation, simultaneity) [41]. Generally, externalized 3D relationships combined with inquiry prompts reduce the need for spatial inferences and increase germane processing focused on principles and constraints.

Next, it should be pointed out that it is very important to work towards optimizing students' cognitive load, within an IVR learning context. In fact, cognitive load may be optimized by signaling, staging, and pre-training [72-73]. Findings attribute learning gains to clear visual signaling, progressive disclosure, and pre-training that orients novices to controls, thereby reducing extraneous load and enabling germane effort [50, 56]. The study from Porter et al. reported that the primary benefit of the training was improved navigation and interaction within the virtual environment. This led to a small but statistically significant improvement in overall electrostatics assessment scores when questions were posed in VR. Training was particularly helpful for non-gamers, helping to bridge the gap in comfort and navigation within the VR space. Additionally, researchers found that VR instruction had no effect for posttest posed on computers [53]. Haptic cues in stargazing appear to aid beginners and, via progressive revelation (a Gollin-like effect), help learners form concrete images from initially abstract visuals [56]. Generally, when interface friction is low and scenes are staged, working memory is reserved for meaning-making, not manipulation.

An IVR feature that is particularly useful for learning about relativity is effortless perspective-taking. Concretely, external projection and shared displays broaden access beyond the headset and support multi-perspective reasoning, for example, toggling between first-person and observer frames in relativity activities [41]. We could assert that perspective switching builds frame awareness and conceptual coherence across representations.

Earlier research has also shown that IVR supports the growth of key cognitive competencies. Students credit IVR activities with developing critical thinking [41, 64] and more expert-like experimental epistemologies (e.g., seeing labs as sites for generating knowledge and models) [36]. Alka et al. describe an IVR Michelson-Morley environment that replicates the experimental setup and lets students align components and optimize fringes, explicitly positioning IVR as a substitute for a delicate apparatus that would otherwise be difficult to use in school settings [32]. Bancong and Nanda, Mufit et al., Šiđanin et al., Wang, and Xie et al. likewise present IVR as a virtual laboratory or experimental space for photoelectric, quantum, nuclear, materials, and circuit activities when real equipment is expensive, hazardous, restricted, or unavailable [8,33,51,63]. Some studies frame IVR not only as a replacement but as a complement to authentic laboratory thinking and procedures. Canright and White Brahmia emphasize experimentation, model development, and realistic measurement tools, while Mufit et al. include tutorials, equations, calculators, and data processing inside the IVR environment so that the full inquiry cycle can occur within or alongside lab-style learning [36, 51]. In addition, gamified IVR can strengthen transferable competencies by structuring purposeful practice [34, 46]. Furthermore, in the study by Kwon participants who learned through the VR system with three-dimensional controllers were able to grasp and manipulate objects in the virtual environment more easily and naturally than those who learned through the VR system with generic gamepad [46]. Overall, our findings indicate that as representational competence rises, students better justify claims with evidence, articulate models, and generalize.

The results from our literature review also suggest that IVR's time efficiency feature allows for reallocating attention to concepts and higher cognitive processes. Concretely, in quantum topics, IVR is rated as highly efficient (e.g., 91% "very practical" on an efficiency indicator), with students perceiving that display design streamlines time [50-51]. Shorter setup and procedural time leave more time for interpretation, cross-representation comparison, and reflection – all drivers of conceptual change. However, efficiency gains are only beneficial if time is redirected into sense-making rather than additional tasks [51].

It is also important to mention that IVR effects may depend on students' characteristics. In fact, students with highly positive attitudes toward science and digital technologies, rate immersive VR as more challenging, complex, and relevant, and evaluate the inquiry process more favorably [58]. In an ORUN-VR context, women rated visuals/realism/interaction higher, while men showed larger gains from control to experimental groups [57]. In a sports-framed mechanics task, male students performed a higher number of successful attempts, which may be linked to greater prior gaming and VR experience as well as differing strategic approaches [61]. In the study by Madden et al., the researchers found differences favoring male students, which were highly correlated with gaming experience. On the other hand, past experience with VR did not improve students' learning in the VR condition [48]. Additionally, Porter et al. discovered that gamers score higher on the pretest than nongamers but on the former set of the questions there was no statistically significant difference between gamers and nongamers [53]. In other words, gaming/VR familiarity can shape performance and motivation patterns; onboarding helps level the field [44, 61].

We also analyzed how IVR helps overcome specific challenges in learning different content areas of physics. Our findings are presented in Table 4.

**Table 4**. Specific challenges of learning different content areas of physics and corresponding IVR affordances

| Content Area | Specific Challenges | IVR Affordances |
|---|---|---|
| Astronomy | Scale, distance, and motion are non-intuitive; many constructs are invisible/abstract; public learners vary widely in prior knowledge. | Free movement and life-like vistas anchor scale reasoning (solar-system sizes, black-hole contexts) and support co-attention via shared displays; facilitation patterns (e.g., active guiding) prompt mechanistic talk rather than passive awe [43]. Haptic/interactive cues and progressive revelation help novices form concrete images from abstract star fields [56]. Learners report some ideas would remain unclear without VR [59]. Participants who preferred VR generally reported that viewing the complete system in a realistic manner enhanced their learning. The primary factors contributing to this perception were the ability to easily observe the system from multiple perspectives and the increased sense of control over the simulation [48]. |
| Electrostatics | Vector fields are invisible; students struggle to connect diagrams, lines of force, and physical interpretation. | Immersive cues scaffold field-line topology and source–field relationships, improving diagram quality and verbal explanation; coordinated views (between the diagram and the 3D field) reinforce representational coherence [31]. |
| Special theory of relativity (STR) | Frames of reference and simultaneity are counter-intuitive; learners conflate representations with reality; difficult to experience consequences directly. | Inquiry-driven scenarios with observer versus traveler vantage points make axioms and consequences (length contraction, time dilation, simultaneity) experiential. External projection enables shared perspective-taking, aligning group reasoning and clarifying frame dependence [41]. |
| Mechanics (projectile motion/trajectories) | Students over-rely on formula recall; struggle to link parameters to motion and to test predictions systematically. | Direct variable manipulation and real-time feedback support hypothesis testing, trajectory reasoning, and iterative refinement, shifting learners from procedural recall to conceptual mastery [61]. |
| Quantum physics | Abstract entities and measurements lack everyday analogues; cognitive load spikes with unfamiliar apparatus/notation. | Experiment-oriented, staged sequences with clear display design increase efficiency and keep attention on model-building; students report more complete information from 3D views and value systematic, theory-informed experimentation [50-51]. |

Finally, we can conclude that across physics topics, several empirical studies reported that IVR may support conceptual understanding [31, 61], representational competence [63], critical and creative thinking [41,64], and inquiry practices through vivid visualizations [58], optimizes cognitive load [12], and coordinates social attention [41, 52]. Concretely, by externalizing 3D structure, motion, and invisible fields – while letting learners manipulate variables and perspectives – IVR reduces the spatial inference burden and helps students coordinate multiple representations such as diagrams, graphs, vector fields, and symbolic forms [31, 42]. Socially, shared views and live facilitation align joint attention, making explanation and comparison easier. It seems the strongest cognitive gains appear when vivid representations are didactically engineered – with signaling, staged complexity, shared views, and facilitation – to channel perceptual richness into model-based reasoning and conceptual change [74-75].

Our model of IVR-based learning of physics is vividly presented in Figure 2. This figure summarizes the review findings by presenting IVR not as a direct cause of learning outcomes, but as a set of affordances whose educational value depends on the cognitive mechanisms they activate under appropriate instructional design and scaffolding. More specifically, the figure suggests that IVR supports physics learning when immersive experience is transformed into disciplined forms of thinking, such as noticing relevant structures, coordinating representations, manipulating variables, and using evidence to refine explanations. Thus, its central implication is that learning gains should be interpreted as emerging from the alignment between IVR affordances, cognitive activity, and physics-specific epistemic work, rather than from immersion alone.

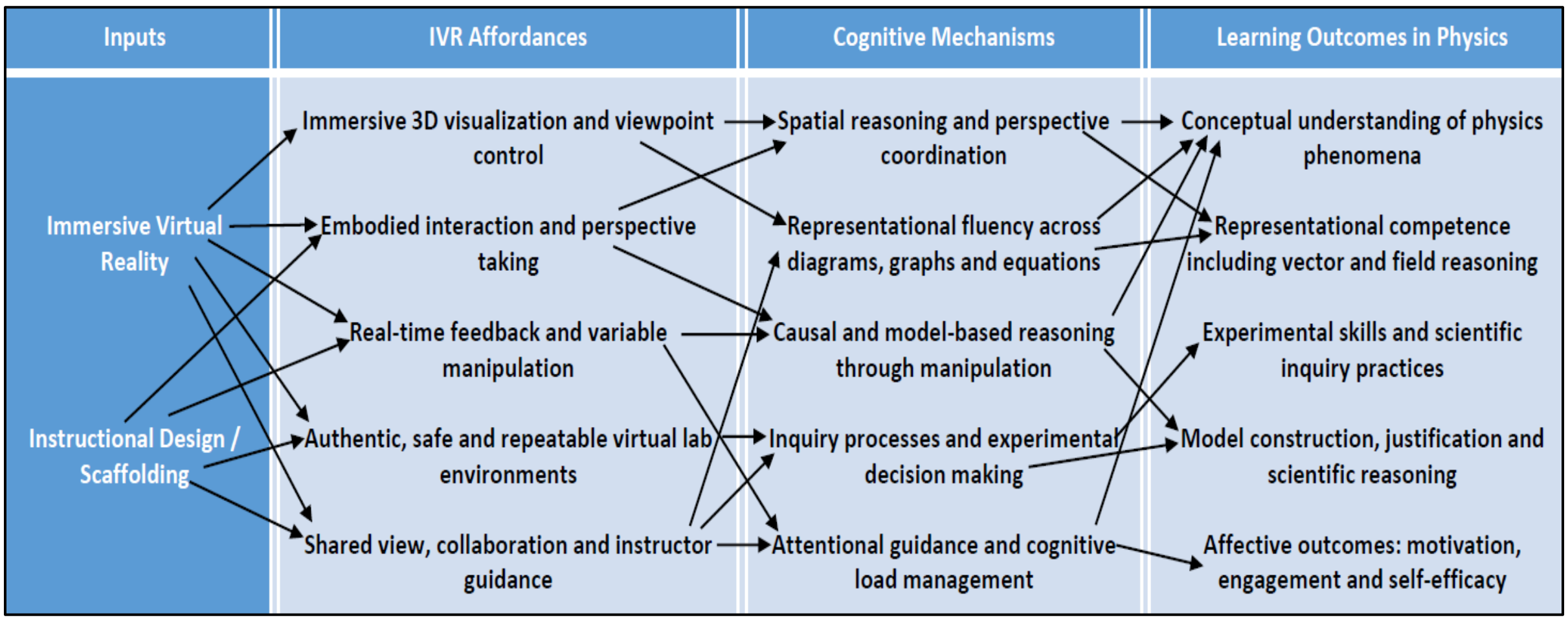

Figure 2: Conceptual synthesis of how IVR affordances support cognitive mechanisms and physics learning outcomes.

Finally, findings from wider literature suggest that cognitive performance is often mediated by some affective variables. Specifically, Makransky and Mayer (2022) found that enjoyment mediated students' performance on an immediate post-test related to an IVR-based climate change learning intervention, whereas interest mediated performance on the delayed post-test [19].

## 3. Ease of use and practicality

For an educational technology to matter system-wide, it must be easy to adopt and practical to run. In classroom reality, every extra click or confusing control taxes extraneous cognitive load and diverts attention from physics. Usable educational technology feels natural, responsive, and "invisible", allowing students to focus on thinking rather than managing the tool. Accessibility then extends that usability to budget, safety, and inclusion: can schools afford it, can it replace or augment risky or scarce apparatus, and can diverse learners participate comfortably? When educational technology is easy, affordable, safe, and inclusive, it creates the preconditions for sustained learning.

Next, we present findings from earlier research on IVR ease of use and practicality. A first important result is that students and visitors frequently describe IVR environments as easy to use, practical, and authentic when controls are responsive and navigation is intuitive [43, 47]. Usability surfaces as a dominant theme in feedback; clarity of instructions and near-invisibility of the technology support the sense that learners are *doing physics* rather than *operating a device* [43, 51]. As a result, when interaction feels natural, extraneous load drops and more working memory becomes available for model-based reasoning.

Another important result is that shared displays help access, but do not fully replace immersion. External projection and shared screens broaden access (letting non-headset peers see and discuss), yet they do not replicate full immersion or resolve the single-user constraint [41]. In other words, shared views are excellent for orchestration and discussion, but headsets remain important for first-person spatial reasoning.

Additional IVR features that are often appreciated are time and cost efficiency. Concretely, learners rate IVR very highly for time efficiency [50, 51]. Furthermore, IVR can reduce wear and tear, enable repeatable practice, and substitute for some expensive or scarce apparatus, improving scalability when budgets are tight [50, 51]. Generally, it should be noted that time and cost efficiencies only contribute to learning when the "saved" capacity is reinvested in interpretation, comparison, and reflection.

Another important feature of IVR technologies is that they allow for safe learning about some potentially risky procedures. Specifically, IVR can simulate hazardous steps (e.g., high-voltage or fragile apparatus) so students rehearse safely. Thereby, guided overlays and prompts can standardize safety-critical actions. Live interaction also reduces fear of asking questions, which is key for safe and reflective practice [47]. Generally, we can say that safety and usability jointly increase willingness to engage, keeping attention on conceptual aims rather than fear or confusion.

In conclusion, evidence from various settings indicates that IVR can be practical, usable, and scalable: students find systems easy to handle, instructors can orchestrate whole-class activity with shared views, and schools can leverage IVR to save time and substitute for some costly or risky apparatus [41, 43, 47, 50, 51]. Usability and accessibility are not mere logistics – they are essential learning conditions. When onboarding lowers frustration, controls feel natural, access is broadened with shared displays, and equity is monitored and supported, IVR becomes a reliable platform for sustained cognitive and motivational gains, not just a one-off novelty[44, 56, 57, 58, 61].

### 3.2.2. Potential disadvantages

Like any educational technology, immersive virtual reality (IVR) carries risks that depend on context, including increased cognitive and logistical demands as well as equity and safety considerations. In the following lines, we outline the main categories of potential disadvantages and the conditions under which they are most likely to arise, followed by practical mitigation strategies.

#### 1. Discomfort and health risks

Extended wear can interfere with attention and shorten productive sessions [47]. In addition, some researchers report blurriness, dizziness, disorientation**,** or motion- related discomfort, often emerging after 15 or more minutes of use [34, 41, 48]. In public/visitor settings, most feedback is positive, but in a study by Kersting et al. participants report disorientation during movement [43]. Discomfort shortens sessions and raises frustration, which is inversely related to flow in IVR contexts [56]. Additionally, Kwon reported that discomfort in using VR systems could occur due to connected cables from computer where cable can wound around students' body as they turned and moved [46].

#### 2. Availability and orchestration overhead

Relatively high acquisition costs typically restrict the number of devices, whereas pragmatic targets like two headsets per classroom limit concurrent immersion [41]. On the other hand, limited headset inventory and single-user constraints create turn-taking overhead: while external projection broadens access, it does not replicate immersion [41]. We can generally say that low device density increases turn-taking and can depress participation equity unless shared displays and roles are carefully designed.

#### 3. IVR interface and cognitive overload

Earlier research has identified potential interface-related issues. Specifically, low-fidelity graphics, timer-based interactions, and raycasting click issues degrade usability [31, 39, 41]. Furthermore, students request explicit concepts (e.g., equipotential surfaces), reset buttons**,** and more reliable UI affordances [31, 39]. In the study by Urhan and Akpinar approximately 50% of students reported no problems, while the remainder needed teacher/researcher assistance with failures such as button contacts and hardware/OS issues that stalled activity [59].

Some researchers identified potential issues related to cognitive load. Concretely, in gameplay-like contexts, off-screen threats or startling "seductive details" can capture attention, eliciting startle, evasive behavior. Additionally, students reported an overwhelming or complex experience. This effect habituates but can still divert cognitive resources from core concepts [48, 57].

Finally, it should be noted that **e**xtra time is often required to learn the equipment (extraneous load) and to process exploratory feedback. Therefore, comparisons to non-IVR groups can be confounded if time budgets are not matched to feedback complexity [63].

In general, poor cueing combined with competing visual events and the requirement to learn about equipment may result in additional extraneous processing and weaker conceptual focus.

In conclusion, the most frequent IVR drawbacks center on discomfort, poor interaction reliability, orchestration issues, cost/availability issues, and extraneous cognitive load. These disadvantages are conditional – they become more pronounced with heavy or poorly optimized hardware, fragile user interface (UI) designs, limited device availability, and surprise-laden scenes under tight time budgets. Conversely, comfort-first setups**,** robust, explicit UIs**,** equity-minded orchestration, seductive-detail control, and well-matched time-on-task can turn many liabilities into manageable constraints [31, 34, 39, 41, 43, 47, 57, 59, 63]. In practice, the question is less **"**Does IVR have disadvantages?" and more "Under what conditions do they arise, and how can design address them?"

# 4. Suggestions for improving IVR-based learning

Our paper reviews how IVR technologies can facilitate physics learning and outlines their potential advantages/disadvantages based on earlier IVR studies. First of all, it is important to note that physics teachers need a combination of technological, pedagogical and content knowledge in order to competently decide in which physics learning contexts using IVR may add value. This particularly holds for topics/procedures that are highly abstract, potentially dangerous, expensive, or hard to realize in a school lab. They also need skills in scaffolding IVR lessons so that students are guided, not overwhelmed. The presented studies repeatedly show the value of embedded tutorials, structured missions, immediate feedback, mirrored displays, collaborative roles, virtual clipboards, quizzes, and on-screen equations or calculators that support the full inquiry cycle [33, 36, 37, 48, 50]. In our view, IVR can be integrated into physics teaching even in large classes. The reviewed studies indicate that it is most effective when instructors are trained to implement it as a shared, scaffolded learning activity. Rather than providing each student with an individual headset—which is more expensive—IVR appears to work better when used as a collaborative learning experience supported by structured guidance from the instructor [43, 47, 52]. What seems uniquely physics-specific about IVR learning effects is that IVR does not merely increase engagement, but supports core forms of physics learning by making invisible phenomena perceptible, linking embodied action to abstract concepts, and enabling real-time coordination of visual, spatial, numerical, and symbolic representations [48, 50, 57]. Because learning physics often includes abstract content,

IVR can function as a bridge between formal theory and perceptual experience, especially when students manipulate variables, observe consequences, and reason from evidence within a structured virtual environment. IVR affordances appear to interact with core physics epistemic practices by turning abstract reasoning into situated, manipulable inquiry: students can build and revise models by changing variables and observing consequences, coordinate vector reasoning through real-time overlays of force, acceleration, or field direction, and develop field-based understanding by inspecting three-dimensional electric, magnetic, or relativistic representations from multiple perspectives [36, 53, 57, 58]. In addition, IVR supports experimental design by allowing learners to set conditions, manipulate apparatus, collect observations, compare outcomes, and draw evidence-based conclusions within simulated environments that approximate authentic laboratory work while remaining safe, repeatable, and accessible [8, 41, 49, 51]. For physics specifically, this is important because the medium helps connect formal representations, embodied action, and empirical reasoning, so that modeling, visualization, and experimentation are not treated as separate activities but as mutually reinforcing ways of knowing in physics. Based on the reviewed studies, IVR may affect conceptual change in physics somewhat differently from many other STEM fields because physics learning so often depends on revising misconceptions about invisible forces, fields, motion, scale, reference frames, and other abstract relationships that cannot be directly observed in ordinary classroom settings. In these studies, IVR supports conceptual change not simply by presenting information more vividly, but by allowing learners to manipulate variables, inspect dynamic three-dimensional representations, and coordinate embodied action with formal concepts such as force, acceleration, equipotentials, field structure, and relativistic effects [31, 36, 41]. That said, in the IVR context, the most important students' activities are: basic VR operation, conceptual grounding in the topic, spatial interpretation, variable control, data recording, and evidence-based reasoning. These activities let students treat IVR as a physics lab for thinking, not just a headset for viewing. In this sense, IVR becomes epistemically meaningful when students use interaction not merely to observe simulations, but to formulate claims about physical systems, compare them with simulated observations or data, and justify them using appropriate models and representations.

Based on the literature review, we suggest the following measures for improving students' outcomes in the affective domain of learning:

- **Onboarding-** short, explicit training for teachers and students supports motivation and satisfaction with achievement [34].
- **Engineer presence for participation-** leverage live audio/avatars and shared views; invite experts; structure think–pair–share activities [47, 52].
- **Design for flow-** calibrate challenge, provide clear goals and feedback, and minimize friction to reduce frustration [56].
- **Plan beyond novelty-** expect early spikes in affective states; sustain them with meaningful tasks, scaffolding, and reflection prompts [36].

For the cognitive domain of learning, the most important general suggestions are as follows:

- **Onboarding** - short, explicit training for teachers and students reduces extraneous cognitive load.
- **Anchor IVR in student-centred activities -** frame sessions around questions, evidence, and model building; use IVR to provide otherwise impossible experiences (e.g., relativistic scenarios) [41, 50]. Roles, quests, or scoring can increase engagement without sacrificing rigor [37, 61].

- **Manage cognitive stimulation -** minimize surprise events; use signaling and progressive disclosure; Keep background stimuli stable during reasoning episodes [57].

**For the specific context of physics learning, we recommend the teachers to:**

- **Prioritize physics-specific abstractions** - use IVR especially for topics that are highly abstract, spatially complex, or difficult to realize in school laboratories, such as electric fields, relativity, particle interactions, decay series, and crystal structures [9, 38, 55, 58, 65].
- **Support vector and field reasoning** - allow students to inspect vectors, field structures, and changing spatial relations from multiple viewpoints in real time, rather than relying only on static diagrams [9, 35, 57].
- **Connect action with concept** - design tasks in which students grab charges, throw objects, move through space, or manipulate apparatus while observing corresponding changes in force, acceleration, trajectory, or equipotential structure [31, 61, 76].
- **Embed representations in the scene -** place measurements, vectors, prompts, equations, and other symbolic supports directly next to the relevant phenomena to strengthen links between observation and interpretation [48, 50, 57].
- **Use IVR for conceptual change** - apply it where everyday intuition conflicts with formal physics concepts, especially in relativity, field behavior, and multidimensional motion [41, 53, 58].
- **Exploit safe repeatability** - use IVR when real experiments are hazardous, expensive, delicate, or inaccessible, as in photoelectric, X-ray, Compton-effect, nuclear, or laser-plasma contexts [8, 33, 49, 50, 51].

Finally, we offer suggestions related to students' health and comfort, as well as to organization of learning activities and IVR interface design:

**a) Comfort-first hardware and session design**

- Use lighter, ventilated HMDs, balance front weight with rear support, and calibrate IPD and straps.
- Keep brief, focused modules with scheduled breaks; prefer seated modes when suitable.
- Limit continuous exposure for novices to 10–15 minutes and consider that longer, movement-heavy tasks raise discomfort risk [34, 43, 47].
- Avoid cardboard-class headsets, small screens, hot rooms, and HMD connected via cables as these degrade comfort and visibility [39, 46].

**b) Onboarding and contingencies**

- Start with a micro-tutorial (movement, interaction, one physics action).

- Prepare fallback options (spare device or alternate activity) in case of hardware failure [59].

**c) Equitable classroom orchestration**

- Pair headsets with shared displays; assign observer and guide roles with structured turn-taking.
- Use shared views for teacher prompts and peer explanation; rotate headset use for equity.
- Note that external projection augments but does not replace immersion [41].

**d) Robust, transparent UI**

- Replace timer triggers with direct actions; improve ray-cast click reliability.
- Make key constructs explicit in the interface to guide attention [31, 39].

**e) Differentiation of instruction**

- Offer varied content delivery by utilizing different scenarios, multiple viewpoints, and alternative display formats to suit diverse learning needs [36, 52].
- Accommodate varying readiness levels through multiple interaction modes, autopilot or simplified controls, embedded tutorials, and guided missions [37, 50, 58, 65].
- Facilitate inclusive participation via accessibility-oriented design choices, mirrored displays for shared participation, and collaborative task structures that allow students to engage in different roles, such as active users or observers [43, 47, 52, 59].

In conclusion, effective IVR instructional design in physics requires careful topic selection, clear student onboarding, inquiry-based tasks, co-located representations, coordinated collaboration, and reflective activities that help students connect the immersive experience with formal physics reasoning. We believe that educators would greatly benefit from a central, high-quality library of physics IVR apps and teaching guides, much like PhET Interactive Simulations.

# 5. Limitations of the study

This systematic review has several limitations. First, the literature search was restricted to Scopus and ERIC. Other major databases, such as Web of Science and Google Scholar, were not searched. Google Scholar was excluded because its broad indexing also includes articles of very poor methodological quality that could bias conclusions about IVR in physics learning. On the other hand, a large majority of articles that are indexed in Web of Science are also included in Scopus, which means that adding Web of Science would most likely not result in a broader article pool.

Second, the review was limited to English-language, peer-reviewed journal articles. Future reviews could broaden the evidence base by incorporating additional high-quality sources, such as conference proceedings, dissertations, theses, and technical reports, and by including studies published in other languages. This is important because publication practices in educational research may favor statistically significant and more positive findings, which can contribute to publication bias. Consequently, the literature available for review may itself overrepresent the beneficial effects of IVR.

A related limitation is that the present synthesis may understate some disadvantages of IVR. Because the review was restricted to physics-specific studies and to the available published journal literature, disadvantages discussed more explicitly in the broader IVR literature may not have been equally represented in our dataset. This is especially relevant for cognitive load: immersive environments can increase extraneous cognitive demands when learners must orient themselves in unfamiliar virtual spaces, master new interaction techniques, and divide attention between conceptual content and interface management. Although pre-training, onboarding, and carefully scaffolded task design may alleviate these challenges, such supports were not implemented uniformly across the reviewed studies. Accordingly, the positive trends identified in this review should be interpreted with caution and understood as dependent on instructional design conditions that help prevent overload and support adaptation to the immersive environment.

Another limitation is that we did not conduct formal intercoder agreement checks for the full pool of screened articles or for the full categorization of IVR advantages and disadvantages. Although many screening and coding decisions were relatively straightforward and low-inference, the absence of full intercoder checks remains a methodological limitation.

In addition, this review did not use a formal quality assessment or risk-of-bias tool for the included studies. Because the synthesis was predominantly qualitative and narrative, we did not systematically grade the methodological strength of each study. Restricting the review to peer-reviewed journal articles indexed in Scopus and ERIC provided a basic quality filter, but it does not substitute for a standardized appraisal of methodological quality. Therefore, the synthesized findings should be interpreted with the understanding that the review cannot explicitly distinguish between stronger and weaker evidence.

Finally, no formal methodological quality assessment or standardized risk-of-bias appraisal was conducted. Consequently, the findings should be interpreted as a structured qualitative synthesis of reported opportunities and challenges rather than as a graded evidence assessment of IVR effectiveness. This limitation is especially important because the included empirical studies differed substantially in research design, participant level, learning context, dependent variables, and outcome measures. Therefore, conclusions about the benefits of IVR should be understood as reported tendencies across the included studies, not as evidence that IVR produces uniformly strong positive effects across physics topics and instructional contexts. The limited number and heterogeneity of empirical studies also made meta-analysis inappropriate.

# 6. Conclusions

This paper outlines the results of a systematic literature review designed to pinpoint opportunities and challenges in using immersive virtual reality (IVR) for physics education. The analysis covered articles from Scopus and ERIC databases, published between January 1, 2018, and June 1, 2025, that specifically addressed IVR applications in physics learning. Furthermore, because this review did not include a formal methodological quality assessment or standardized risk-of-bias appraisal, the conclusions below should be read as a synthesis of reported patterns and design possibilities rather than as a graded evaluation of the strength of causal evidence. Accordingly, claims about IVR benefits are framed as potential affordances and tendencies reported in the included literature.

Within the limits of the available evidence, the reviewed studies suggest the following potential IVR affordances related to the cognitive domain of learning:

- **Making unobservable and abstract phenomena experientially accessible** through real-time visualizations that reduce extraneous cognitive load. This allows learners to immediately connect variable manipulations with physical outcomes (e.g., observing Lorentz contraction at relativistic speeds, visualizing 3D electric field configurations, and tracking simulated particle interactions) [9, 35, 38, 41, 43].
- **Replacing unavailable, expensive, or hazardous equipment** by simulating experiments that require delicate or restricted setups, such as interferometers, nuclear detectors, photoelectric apparatuses, or advanced laser–plasma systems [8, 32, 33, 49].
- **Supporting embodied cognition** by linking bodily movement and hand-based interactions with abstract physics ideas, enabling learners to grasp, throw, or navigate virtual objects while observing corresponding real-time changes in forces, acceleration, and spatial relationships [31, 61, 76].
- **Facilitating the experimental method and causal reasoning** by allowing learners to formulate hypotheses, manipulate system variables (e.g., projectile kinematics or friction coefficients), and collect measurements within a highly controlled simulated laboratory [32, 36, 50, 57, 64].
- **Providing repeatable, high-fidelity practice** for complex, microscopic, or probabilistic physical processes, such as simulating X-ray production or the Compton effect [50, 51].
- **Granting absolute control over spatiotemporal scales**, such as accelerating celestial motion or enabling astronomical observations across changing dates and geographical locations [52].
- **Optimizing collaborative orchestration and instructional guidance** by embedding real-time signaling, annotations, and shared environments for peer discussion directly within the physics simulation [35, 47].

Besides potential benefits for the cognitive domain of learning, IVR has the potential to positively impact the affective domain of students' learning. Concretely, IVR can:

- increase motivation, interest, and engagement through game-like elements, missions, rewards, role-play, and narrative framing [37, 51, 58]
- enhance presence, attention, and self-efficacy by placing learners "inside" the phenomenon or lab, often with 360° environments and responsive graphics [43, 47, 57]
- stimulate interest in physics and STEM careers by virtually visiting real research labs and "seeing through the researcher's eyes" [38, 51]

Finally, IVR may also significantly facilitate the organization of learning in contexts that are hard to access. Specifically, it has the potential to:

- support the full scientific inquiry cycle within the IVR environment (hypothesis formulation, experimentation, data collection, analysis, and conclusion) by embedding comprehensive learning tools such as tutorials, quizzes, virtual calculators, notebooks, and computers [32,33]
- improve accessibility for outreach and distance education (e.g., festivals, museums, remote SDO solar data) while still providing rich, interactive learning experiences [43, 47, 56]

When it comes to the most frequent IVR challenges, they mostly center on discomfort, interaction reliability, orchestration challenges, cost/availability constraints, and extraneous cognitive load. In fact, recent research suggests that the strong sense of presence that makes virtual reality highly engaging can also increase cognitive load and thereby hinder learning. When users are deeply immersed perceptually and psychologically, the immersive qualities of VR may interfere with effective knowledge acquisition. One potential solution is the dynamic detection of cognitive load, which would allow the level of presence in the virtual environment to be adjusted to support optimal learning outcomes [15]. Mentioned disadvantages intensify with poorly tuned hardware, fragile UIs, limited devices, and surprise-laden scenes under tight time constraints. Conversely, comfort-focused designs, robust interfaces, equity-minded orchestration, seductive-detail control, and appropriate time-on-task can convert many liabilities into manageable constraints.

It is very important to note that whether IVR will benefit learning or make it more difficult largely depends on the interaction of IVR potential, learning opportunities created by the teacher and by students' characteristics. Finally, this literature review also reveals several important research gaps. In fact, most studies on IVR in physics are short-term, so long-term retention and sustained learning outcomes are still not well understood. More evidence is also needed from authentic classroom settings, including large classes, environments with limited equipment, and routine school implementation. The reviewed research further indicates that teacher orchestration, cognitive load, learner comfort, and equitable participation are important factors in IVR-supported learning. These aspects should be further investigated in future studies.